%% file: cer-speeches.tex
\documentclass[12pt]{article}

\usepackage[a4paper,pdftex]{geometry}
\usepackage[T1]{fontenc}
\usepackage[english]{babel}
\usepackage[authoryear]{natbib}
\usepackage[small]{titlesec}
\usepackage[justification=centering]{caption}
\usepackage{hyperref}

\usepackage{amsmath,amsthm,amssymb,graphicx,enumerate,booktabs,bigstrut,rotating,multirow,float,etoolbox,lmodern,comment,listings,qtree,a4wide,moresize,bbm,subcaption,tikz,pdfescape,mathtools,flafter,enumitem,setspace,ragged2e,colortbl,lineno,lscape,pdflscape,stackengine}

\newcolumntype{L}[1]{>{\raggedright\let\newline\\\arraybackslash\hspace{0pt}}m{#1}}
\newcolumntype{C}[1]{>{\centering\let\newline\\\arraybackslash\hspace{0pt}}m{#1}}
\newcolumntype{R}[1]{>{\raggedleft\let\newline\\\arraybackslash\hspace{0pt}}m{#1}}

\hypersetup{colorlinks,linkcolor={red},citecolor={blue},urlcolor={blue}}

\makeatletter
\renewcommand\subsubsection{\@startsection{subsubsection}{3}{\z@}%
	{-3.25ex\@plus -1ex \@minus -.2ex}%
	{-1.5ex \@plus -.2ex}%
	{\normalfont\normalsize\bfseries}
}
\def\@biblabel#1{\hspace*{-\labelsep}}

\makeatother

\def\sym#1{\ifmmode^{#1}\else\(^{#1}\)\fi}

\pdfpageheight\paperheight
\pdfpagewidth\paperwidth

\begin{document}
	
%	\title{The Shifting Attention of Political Leaders: Evidence from Two Centuries of Presidential Speeches}
	
	\title{The Shifting Attention of Political Leaders: Evidence from Two Centuries of Presidential Speeches\thanks{\noindent Calvo-Gonz\'alez: The World Bank (ocalvogonzalez@worldbank.org), Eizmendi: Tufts University (axel.eizmendi\_larrinaga@tufts.edu), Reyes: briq Institute on Behavior \& Inequality (german.reyes@briq-institute.org). For helpful comments, we thank Aviv Caspi, Laura Chioda, Alexandru Cojocaru, Christa Deneault, Bill Maloney, Ambar Narayan, Roy Van der Weide, Daniel Valderrama, and participants from the Poverty GP lunch seminar. We also thank Evangelina Cabrera, Martha Narvaez, Sandra Pastr\'an, Cesar Polit, Yalile Uarac, for compiling and forwarding to us multiple speeches from Argentina, Chile, Ecuador, and Paraguay. This research was funded by the World Bank Research Department. Errors and omissions are our own.}} 
	
	\author{Oscar Calvo-Gonz\'alez \and Axel Eizmendi \and Germ\'an Reyes}
	
	\renewcommand{\today}{\ifcase \month \or January\or February\or March\or %
		April\or May \or June\or July\or August\or September\or October\or November\or %
		December\fi \ \number \year} 	
	
	\date{\vspace{15pt} \today}
	
	\maketitle
	
	\begin{abstract}
		
		\noindent We use natural-language-processing algorithms on a novel dataset of over 900 presidential speeches from ten Latin American countries spanning two centuries to study the dynamics and determinants of presidential policy priorities. We show that most speech content can be characterized by a compact set of policy issues whose relative composition exhibited slow yet substantial shifts over 1819-2022. Presidential attention initially centered on military interventions and the development of state capacity. Attention gradually evolved towards building physical capital through investments in infrastructure and public services and finally turned towards building human capital through investments in education, health, and social safety nets. We characterize the way in which president-level characteristics, like age and gender, predict the main policy issues. Our findings offer novel insights into the dynamics of presidential attention and the factors that shape it, expanding our understanding of political agenda-setting.
		
		%JEL codes: D78, I38, H50
		
		%Keywords: Political leaders, issue attention, natural-language-processing algorithms, machine learning, Latent Dirichlet Allocation.
		
	\end{abstract}
	
	\clearpage
	
	\section{Introduction} \label{sec:intro}
	
	Issue attention---the policy issues that political actors pay attention to---is a key input to models of policymaking. For example, according to agenda-setting theories, issue attention is required to generate policy changes \cite[e.g.,][]{kingdon1984agendas,carmines1986structure,baumgartner2010agendas}. Similarly, the temporal patterns of issue attention are crucial for understanding the timing of future policy changes \citep{downs1972up, peters1985search, cairney2019understanding}.
	
	Recent work uses natural-language-processing (NLP) algorithms to study political attention \citep{quinn2010analyze,grimmer2013text}. By analyzing vast quantities of text, these methods can uncover the main issues discussed, under the assumption that issue attention is revealed by the relative allocation of expressed content. Automatized statistical algorithms have been used to study the content of treaties \citep{spirling2012us}, political e-mails \citep{mathur2023manipulative}, legislators' tweets \citep{barbera_2019}, Federal Open Market Committee meetings \citep{hansen2018transparency, caspi2020measuring}, and congressional speeches \citep[e.g.,][]{herzog2015most, goet2019measuring, osnabrugge2021playing}.
	
	Yet, there has been little work using automatized methods to measure the political attention of presidents. The lack of such a study is significant given the central role of the presidency in shaping and directing policy. In this paper, we apply NLP methods to a novel dataset of presidential speeches to uncover the main expressed policy priorities of presidents, study their determinants, and analyze how they evolve over time.
	
	We use a hand-collected dataset of over 900 annual presidential ``state-of-the-union''-type speeches spanning ten Latin American countries. In these speeches, presidents provide an overview of policies undertaken by their administration and reflect on the priorities for the upcoming years. Our dataset dates as far back as 1819, enabling us to examine presidential discourse throughout significant historical periods. Among other events, our dataset covers a wide range of military conflicts, starting with the independence wars in which Latin American countries gained autonomy from Europe and covering both World Wars; multiple economic crises, including both the Great Depression and the Great Recession; and the rise to power of extremist leaders, both in the far right in the form of military dictatorships and the far left in the form of populist regimes.	
	
	To recover the policy issues discussed in the presidential speeches, we use a natural-language-processing algorithm called Latent Dirichlet Allocation (LDA). LDA uses the words in a set of documents as the only observable variables \citep{blei2003latent, blei2006dynamic, blei2012probabilistic}. An attractive property of LDA is that it does not require the researcher to specify a set of topics into which the documents are classified.\footnote{LDA is an increasingly-popular tool in economics. Researchers typically use it to recover latent types from high-dimensional text data. For example, LDA has been used to measure communication in deliberative bodies \citep{hansen2018transparency}, to study how CEO type affects firm performance \citep{bandiera2020ceo}, and to measure the value of firm amenities covered by collective bargaining agreements \citep{lagos2019labor}.} LDA partitions the dataset of presidential speeches into a set of mutually exclusive and collectively exhaustive ``topics.'' A topic is defined by a probability distribution over the keywords contained in the dataset of speeches. LDA also generates the probability distribution of topics of a given president's speech, which can be interpreted as the proportion of a president's speech discussing each topic. This measure is often used in the literature as a proxy for issue attention.
		
	We use the topics uncovered by LDA for three purposes. First, we show that, despite the high dimensionality of presidential speeches, the expressed policy priorities embedded in the speeches can be characterized by a compact and easy-to-interpret set of issues. Most speech content falls into one of six topics: (i) military conflict and patriotism; (ii) the state of the public administration; (iii) investments in infrastructure; (iv) freedom, individual rights, and social justice; (v) economic development; and (vi) social protection. Across countries and years, these six policy issues together account, on average, for approximately 80\% of presidential speech content.
	
	Second, we study the president-level correlates of expressed presidential priorities. We find that female presidents, older presidents, and democratically-elected presidents are, on average, more likely to discuss economic development and social protection and less likely to discuss war/patriotism and the state of the public administration. Female leaders are also less likely to discuss issues related to infrastructure than male presidents. These findings are consistent with previous work relating political actors' traits to their expressed priorities \citep[e.g.,][]{gennaro_2022, osnabrugge2021cross} or their constituents' outcomes \citep[e.g.,][]{schubert1988age, chattopadhyay2004women, clots2012female}, 
		
	Third, we investigate the dynamics of presidential priorities. We find that the topics discussed in speeches slowly shift over long periods of time that stretch across electoral cycles. Over the course of two centuries, presidential attention gradually shifted from military interventions and the development of state capacity, to building physical capital through investments in infrastructure and public services, and finally to promoting human capital through investments in education, health, and social safety nets. These findings are consistent with issue attention theories positing that policy priorities often exhibit continuity and long-term trajectories due to the enduring nature of societal challenges and institutional constraints \citep{baumgartner2010agendas}.		

	The rest of the paper proceeds as follows. Section \ref{sec:data} describes the data. Section \ref{sec:topic-models} describes topic models and characterizes the issues discussed in presidential speeches. Section \ref{sec:dyn} studies the correlates and dynamics of speech content. Section \ref{sec:conclusions} concludes.

	\section{Data} \label{sec:data}
	
	We assembled a novel dataset consisting of 933 presidential speeches delivered between 1819 and 2021 in ten Spanish-speaking Latin American countries: Argentina, Chile, Colombia, Costa Rica, The Dominican Republic, Ecuador, Mexico, Paraguay, Peru, and Venezuela. These constitutionally-required annual addresses serve as the closest parallel to the United States' ``State of the Union'' speech in these countries. In these speeches, presidents provide an overview of the work performed by their administration and an outline of the policy goals and priorities for the upcoming years. 
	
	The dataset compilation consisted of a two-stage process: collecting the speeches and processing them. We obtained the majority of speeches from Argentina, Ecuador, and Paraguay through their respective National Congresses and librarians. In the case of Chile, Colombia, Costa Rica, the Dominican Republic, and Mexico, speeches were collected through a variety of online sources. Most of Venezuela's speeches were scanned from books available at the US Library of Congress. 
	
	The second stage involved processing the speeches to enable text analysis. For the speeches already in a digitized format, this meant converting each file into text format and removing any text that did not form part of the speech, such as the title or date. For scanned speeches---i.e., those in ``image'' format---we performed Optical Character Recognition (OCR) to convert the images into machine-encoded text. To ensure the quality of the data, we manually reviewed all the OCR-generated text and corrected any inaccuracies. The text analysis was performed in Spanish. English translations shown throughout the paper were done by the authors.
	
	Due to variability in data availability, both online and in the Congressional libraries, not every country nor every decade is represented equally in the dataset (see Appendix \ref{app:data}). Costa Rica, Mexico, Peru, and Venezuela are the most represented countries in the dataset, each with 15--18\% of all speeches. Argentina, Chile, Ecuador, and Paraguay have a moderate representation, each with 7\%--10\% of all speeches. Colombia and the Dominican Republic are equally underrepresented in the dataset, each with only 2\% of total speeches. The majority of speeches (68\%) correspond to 1920--2021, with the remaining 32\% dating back to 1819--1919.
	
	We complement the presidential speeches with data on president demographic characteristics and political regime type. We obtain data on presidential terms and presidents' demographic characteristics (age and gender) from Archigos \citep{goemans2009introducing}. To classify governments as autocratic or democratic, we use data from the Polity 5 project \citep{marshall2020}. We define presidents as democratically elected if their ``polity score'' is positive \citep{persson2009democratic}.

	Table \ref{tab:summ} provides summary statistics on the dataset. Most presidents in our sample are males (97.6\%). Presidents were, on average, 54.6 years old when they delivered their speeches. There is substantial variation in regime type in the dataset, with about half of the speeches corresponding to democratically-elected presidents (51.6\%) and the other half to autocratic ones (48.4\%). The party of almost half the presidents (45.8\%) controlled both legislative chambers when the presidents delivered their speeches.
	
	\begin{table}[H]\caption{Summary statistics on the sample of presidential speeches}
		\vspace{-15pt}
		\label{tab:summ}
		{\footnotesize
			\begin{centering} 
				\protect
				\begin{tabular}{lccc}
					\addlinespace \addlinespace \midrule			
					& Mean & SD  & N   \\
					& (1) & (2)  & (3)   \\
					\midrule 	
					\multicolumn{1}{l}{\hspace{-1em} \textbf{Panel A. Characteristics of the speeches}} \\ 
					\input{results/summ-wor.tex} \midrule
					
					\multicolumn{1}{l}{\hspace{-1em} \textbf{Panel B. Characteristics of the presidents}} \\ 
					\input{results/summ-lea.tex} \midrule
					
				\end{tabular}
				\par\end{centering}
			
			\singlespacing\justify\footnotesize
			\textit{Notes:} This table shows summary statistics on our dataset. President characteristics come from Archigos and are typically available since 1870. Regime type (democracy/autocracy) comes from Polity 5 and is available since 1820. Party control of the Congress and the Senate come from the Database of Political Institutions (DPI) and is available since 1975.

		}
	\end{table}

	\section{The Expressed Policy Priorities in Presidential Speeches} \label{sec:topic-models}
	
	This section describes topic models and examines the main policy issues discussed by presidents in our dataset of speeches.
	
	\subsection{Topic Models and Latent Dirichlet Allocation} \label{sub:topic-models}
	
	Topic models are statistical models used to extract the main themes contained in large, unstructured collections of documents \citep{blei2012probabilistic}. The most widely used topic model is the Latent Dirichlet Allocation (LDA) algorithm \citep{blei2003latent}. Given a collection of documents, LDA discovers the primary topics in each document and the degree to which each document exhibits those topics. LDA assumes that documents are random probability distributions over topics, where a topic is a probability distribution over the documents' words (see Appendix \ref{app:lda-summ}). The only object that is exogenously defined by the researcher is the number of topics to be discovered. However, there are procedures to optimally select this figure, such as perplexity minimization (see Appendix \ref{app:perplexity}).
	
	We use LDA to partition each presidential speech into a set of mutually exclusive and collectively exhaustive topics. To choose the number of topics, we follow the perplexity-minimization criterion, which yields 25 topics (Appendix Figure \ref{fig:perplexity}). For each topic, LDA produces a vector of keywords and the likelihood of each keyword belonging to the topic. We assign labels to the topics based on the most probable words to ease interpretability. LDA also generates the probability distribution of topics in each speech, which we interpret as the proportion of a speech discussing each topic.
	
	\subsection{The Policy Priorities Embedded in Presidential Speeches} \label{sub:issues}
	
	Despite the complexity and high-dimensionality of presidential speeches, we find that a small set of policy issues can be used to characterize most speech content. There are only six topics that exceed 20\% of speech content in at least one decade. Table \ref{tab:topics-speeches} lists and defines these six topics. The labels we assign to these topics based on their most representative terms are: (i) War and patriotism, (ii) Public administration, (iii) Infrastructure, (iv) Rights, freedom, and social justice, (v) Economic development, and (vi) Social protection. On average, these six issues jointly cover 77.1\% of speech content across countries and years. Appendix Table \ref{tab:topics-speeches-t} shows that the main topics discovered by LDA are remarkably similar when varying the number of topics to be discovered by the algorithm.
	
	\begin{table}[H]{\footnotesize
			\begin{centering} 
				\protect\caption{Top ten keywords defining the main topics of presidential speeches} \label{tab:topics-speeches}
				\newcommand\w{1.5}
				\begin{tabular}{lccccccc}
					\addlinespace \addlinespace \midrule
					& \multicolumn{6}{c}{Topic name} \\\cmidrule{2-7}				
					& War and    & Public         &                & Rights and  & Economic & Social \\					
					& Patriotism & administration & Infrastructure & freedom     & development & protection \\									
					& (1) & (2) & (3) & (4) & (5) & (6) \\
					\midrule 	
					\input{results/topics-allspeeches.tex} \midrule
					
				\end{tabular}
				\par\end{centering}
			
			\singlespacing\justify
			\textit{Notes}: This table lists the top ten keywords that define the topics in Figure \ref{fig:topics-speeches}. We only list topics whose probability exceeds 20\% in at least one decade. Each column header shows the manually-assigned label of each topic. We recover the topics and their probability distributions using a Latent Dirichlet Allocation (LDA) algorithm (see Section \ref{sub:topic-models} and Appendix \ref{app:lda}).
			
		}
	\end{table}

\begin{figure}[H]
		\caption{Word clouds of the keywords that define main topics}\label{fig:word-cloud}
		\centering
		\begin{subfigure}[t]{.40\textwidth}
			\caption*{Panel A. War and patriotism}
			\centering
			\includegraphics[width=.7\linewidth]{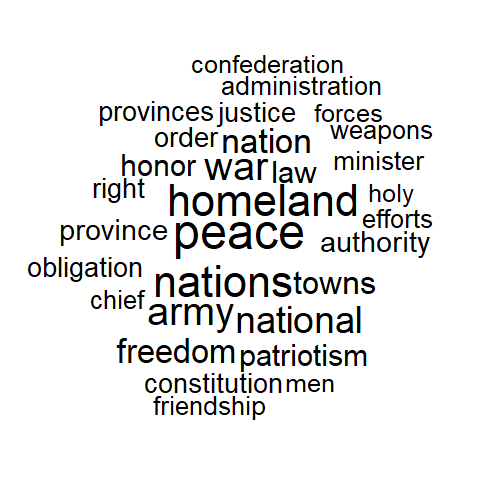}
		\end{subfigure}
		\hfill		
		\begin{subfigure}[t]{0.40\textwidth}
			\caption*{Panel B. Public administration}
			\centering
			\includegraphics[width=.6\linewidth]{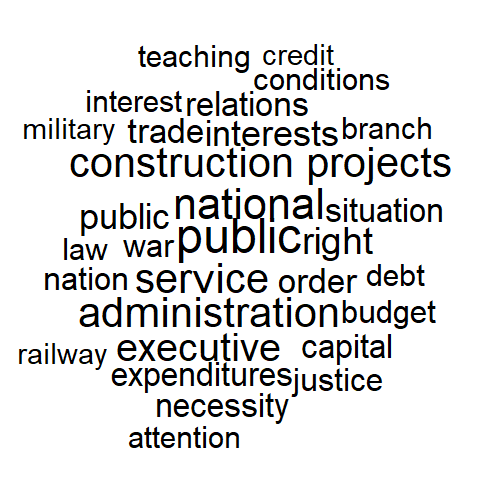}
		\end{subfigure}	
		\hfill		
		\begin{subfigure}[t]{.40\textwidth}
			\caption*{Panel C. Infrastructure}
			\centering
			\includegraphics[width=.6\linewidth]{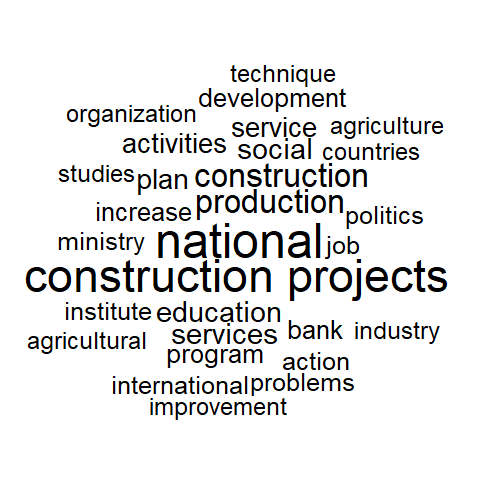}
		\end{subfigure}
		\hfill		
		\begin{subfigure}[t]{0.40\textwidth}
			\caption*{Panel D. Rights, freedom, and social justice}
			\centering
			\includegraphics[width=.6\linewidth]{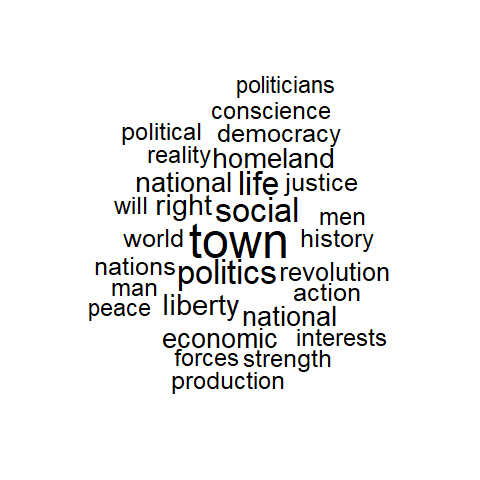}
		\end{subfigure}	
		\hfill		
		\begin{subfigure}[t]{.40\textwidth}
			\caption*{Panel E. Economic development}
			\centering
			\includegraphics[width=.6\linewidth]{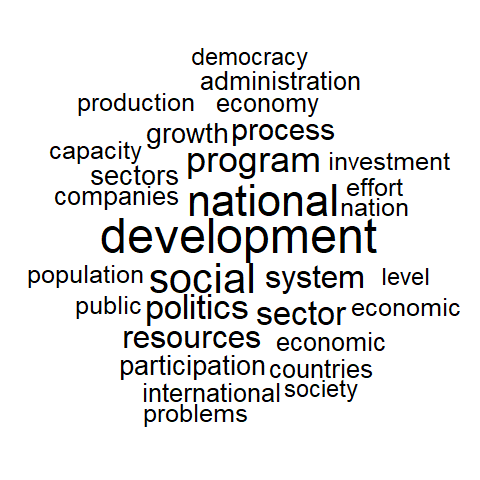}
		\end{subfigure}
		\hfill		
		\begin{subfigure}[t]{.40\textwidth}
			\caption*{Panel F. Social protection}
			\centering
			\includegraphics[width=.6\linewidth]{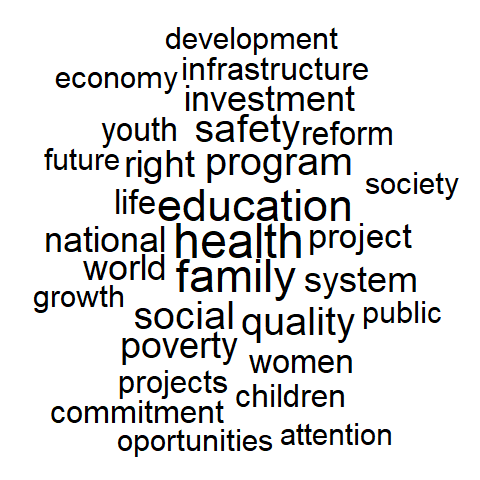}
		\end{subfigure}
		{\footnotesize
			\singlespacing \justify
			\thispagestyle{empty}
			\textit{Notes:} This figure plots word clouds of the distribution of words that define each topic. The size of each keyword is proportional to the importance of the keyword in defining the topic. We estimate the topics and their probability distributions using a Latent Dirichlet Allocation (LDA) algorithm (see Appendix \ref{app:lda}). %We present only topics whose probability exceeds 20\% in at least one decade. Topics are defined by their top occurring keywords (see Table  \ref{tab:topics-periods} for the top ten keywords that define the topics in the figure). We manually labeled topics based on the top keywords. 
			
		}
	\end{figure}

	Figure \ref{fig:word-cloud} displays a series of word clouds plotting the top-defining keywords of each topic. Appendix Table \ref{tab:excerpts} illustrates the content of each topic by providing excerpts from presidential speeches in which LDA assigns a high probability of belonging to a given topic. We discuss in more detail the content of each topic in the context of how the prevalence of these topics in presidential speeches evolved over time. 
	
	% Here we could compare our topics (the # and content) with those of the paper that studies SOTU in the US.

	\section{The Determinants and Dynamics of Policy Priorities} \label{sec:dyn}
		
	This section examines the correlates and evolution of the main policy issues discussed by presidents.
	
	\subsection{The President-level Correlates of Expressed Political Priorities} \label{sub:issue-determ}
	
	We begin by analyzing whether president characteristics correlate with expressed priorities. In Table \ref{tab:det-issue}, we estimate bivariate regressions of the form: 
	\begin{align}\label{eq:det}
		\text{ShareTopic}^k_{it} = \alpha^k + \gamma^k X_{it} + \varepsilon_{it},
	\end{align} 
	where $\text{ShareTopic}^k_{it}$ is the topic $k$ proportion in the presidential speech of country $i$ in year $t$ and $X_{it}$ is a president-level characteristic, such as age, gender, and political regime type (democracy/autocracy). We estimate equation \eqref{eq:det} separately for each of the six main policy issues discussed in the presidential speeches.
	
	President characteristics strongly correlate with expressed priorities (Table \ref{tab:det-issue}). Female presidents are 25.1 percentage points more likely to discuss social protection ($p < 0.01$), 6.0 percentage points more likely to discuss economic development ($p < 0.05$), 21.5 percentage points less likely to discuss public administration affairs ($p < 0.01$), 9.9 percentage points less likely to discuss infrastructure ($p < 0.01$), and 3.4 percentage points less likely to discuss war and patriotism ($p < 0.01$) than male presidents (columns 1, 2, 3, and 6).  
	
	Democratically-elected presidents are 16.7 percentage points more likely to discuss social protection ($p < 0.01$), 12.2 percentage points more likely to discuss economic development ($p < 0.01$), 17.1 less likely to discuss the state of the public administration, and 9.3 percentage points less likely to discuss war and patriotism ($p < 0.01$) than autocratic presidents (columns 1, 2, 5, and 6). President age follows the same qualitative patterns of expressed priorities as democratically-elected presidents. 
	
	In Appendix Table \ref{tab:det-issue-rob}, we show that these patterns are quantitatively similar and qualitatively identical if all characteristics are simultaneously included in the regression equation.

	\begin{landscape}
		%\begin{sidewaystable}
		\begin{table}[H]{\footnotesize			
				\begin{center}
					\caption{The president-level correlates of expressed policy priorities} \label{tab:det-issue}
					\newcommand\w{2.3}
					\begin{tabular}{l@{}lR{\w cm}@{}L{0.45cm}R{\w cm}@{}L{0.45cm}R{\w cm}@{}L{0.45cm}R{\w cm}@{}L{0.45cm}R{\w cm}@{}L{0.45cm}R{\w cm}@{}L{0.45cm}R{\w cm}@{}L{0.45cm}}
						\midrule
						&& \multicolumn{12}{c}{Dependent variable: Fraction of a speech discussing...} \\ \cmidrule{2-14}
						&& War and    && Public          && Infrastructure    && Rights and  && Economic && Social   \\					
						&& Patriotism && administration  &&  && freedom     && development   && protection \\				
						&& (1)        && (2)             && (3)       && (4)         &&   (5)  &&   (6)      \\
						\midrule
						
						\input{results/issue-det-democracy}
						\input{results/issue-det-age}       
						\input{results/issue-det-female} \midrule
						\input{results/issue-det-n}\midrule 		
					\end{tabular}
				\end{center}
				\begin{singlespace}  \vspace{-.7cm}
					\noindent \justify \textit{Notes:} This table presents OLS coefficients, $\gamma^k$, from equation \eqref{eq:det} Each cell displays coefficients from a bivariate regression of the variable listed in the column header on the variable listed in the row header. Heteroskedasticity-robust standard errors in parentheses. $^{***}$, $^{**}$, and $^{*}$ denote significance at the 1\%, 5\% and 10\% levels, respectively.
				\end{singlespace} 	
			}
		\end{table}
		%\end{sidewaystable}
	\end{landscape}	
	
	The positive relationship between female leadership and content devoted to social protection---a topic that stresses the importance of investments in education and health---is consistent with the findings of \cite{clots2012female}, who shows that female politicians increase educational attainment in urban areas in India. The finding that democratic regimes are more likely to discuss economic development is consistent with the literature that links democracy to economic growth \citep[e.g.,][]{acemoglu2019democracy}. Similarly, the link between democratic leadership and expressed attention to social protection is in line with findings showing a positive link between democratic regimes and health outcomes, such as life expectancy \citep[e.g.,][]{besley2006health}.
	
	\subsection{The Evolution of Expressed Priorities Over Time} \label{sub:issue-cycle}
	
\begin{figure}[H]
		\caption{The evolution of expressed presidential priorities over time}\label{fig:topics-speeches}
		\centering
		\includegraphics[width=.85\linewidth]{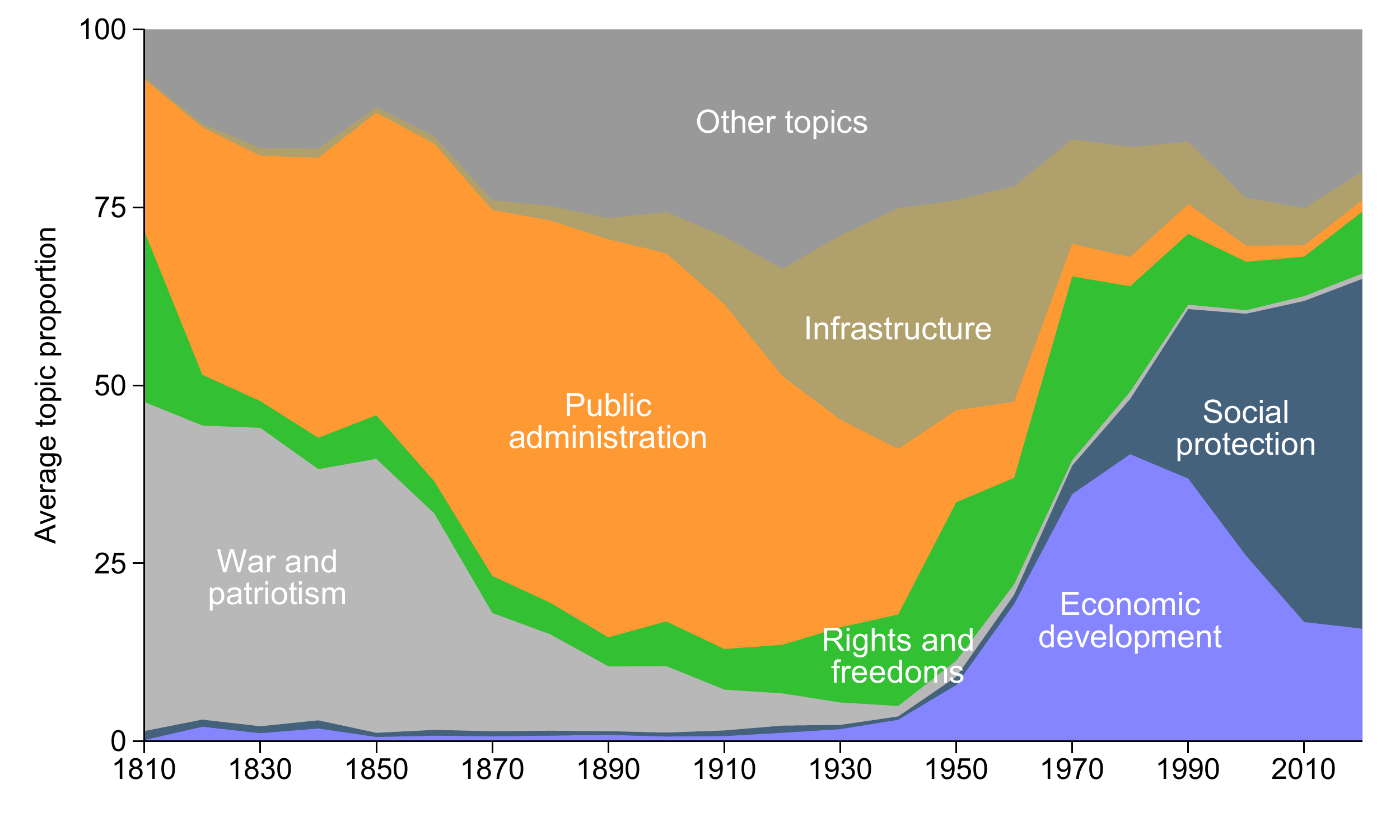}
		\footnotesize
		\singlespacing \justify \textit{Notes:} This figure shows the distribution of topics across decades. We estimate the topics and their probability distributions using a Latent Dirichlet Allocation (LDA) algorithm (see Section \ref{sub:topic-models} and Appendix \ref{app:lda}). To choose the number of topics, we follow a criterion of perplexity minimization (see Appendix \ref{app:perplexity}). We present only topics whose probability exceeds 20\% in at least one decade. The rest of the topics are grouped together in the category labeled as ``Other topics.'' Topics are defined by their top occurring keywords (see Table  \ref{tab:topics-speeches} for the top ten keywords that define the topics in the figure). We manually labeled topics based on the top keywords. To construct the figure, we pool speeches at the decade level and compute the average topic probabilities in each decade. 
	\end{figure}

	Next, we turn to the dynamics of expressed priorities. Figure \ref{fig:topics-speeches} plots the distribution of the main issues discussed in presidential speeches in each decade from 1819--2021. To construct this figure, we aggregate the topics discussed in speeches at the decade level and calculate the average topic proportion in each decade. This enables us to focus on the main topics discussed in each decade but at the cost of ignoring the idiosyncratic year-to-year variation in presidential discourse (captured in our residual ``Other Topics'' category). 
	
	The main topics discussed by presidents tend to change slowly over multiple decades that stretch across electoral cycles. Throughout the 19th century, presidents were primarily concerned with two topics: war/patriotism and the state of the public administration. These two topics combined accounted for roughly 70\% of presidential speech content until the 1870s. The prevalence of the war and patriotism topic is consistent with the historical context of the period, which was characterized by internal conflicts and repeated border wars between the newborn nation-states of Latin America \citep{clayton2017}. The public administration topic captures a range of governance issues, including the organization of government establishments, the provision of public services, and the management of public finances. This likely reflects the fact that the governments and their institutions were still in their infancy, making the development of political order and state capacity a central policy concern \citep{clayton2017}. 
	
	Discussions of war and patriotism decreased after the 1870s. Although the topic of public administration continued to dominate well into the 20th century, beginning in the 1900s, presidents increasingly shifted their rhetoric towards the development of infrastructure and the provision of public services. These policy concerns were particularly dominant during 1920--1950, with attention to this topic reaching its peak in the 1940s. This is consistent with the growing disillusionment with export-led growth during the period and a subsequent shift towards import-substitution industrialization strategies, which emphasized the growth of internal markets, the development of infrastructure, and greater government intervention \citep{prebisch1962economic,bulmer2003economic}. Around this time, presidents increasingly began to discuss a broad topic related to rights, freedom, and social justice, particularly from 1940--1980. Discussions about these issues are perhaps expected, given that this historical period was characterized by the increasing popularity of communist and leftist ideologies---as evidenced by the rise of revolutionary movements across the region---and the rise and fall of military dictatorships, often running on anti-communist agendas \citep{davila2013dictatorship}.
	
	Beginning in the 1950s, there is a stark increase in attention to economic development. During this period, presidents framed economic development as a byproduct of economic growth and changes in a country's productive structure. Accordingly, this topic captures discussions related to economic planning and the management of industrial sectors, which is consistent with the increased adoption of import-substitution industrialization development strategies \citep{bulmer2003economic}.
	
	In the 1990s---a period characterized by the implementation of so-called Washington Consensus policies in several countries of the region \citep{williamson1993democracy, gore2000rise} and a wave of democratization \citep{hagopian2005third}---presidential discourse became increasingly concentrated on issues related to social protection. As the top keywords illustrate, this topic focuses on the importance of building human capital through investments in education and health and stresses the state's role in providing social insurance and building social safety nets. The appearance of the terms ``right,'' ``investment,'' and ``poverty,'' suggests an increasing recognition that access to these public services constitutes a fundamental right that requires government investment and that education and health care are pathways to reducing poverty.\footnote{Another top keyword of this topic is ``programs.'' This is consistent with the rise of large-scale social programs throughout the region, such as conditional cash transfers and non-contributory pensions \citep{ferreira2010social}.} The high prevalence of this topic continued in the 2000s, a period characterized by the surge of leftist governments across Latin America.\footnote{This movement, which is often considered to have started with Chavez's accession to power in Venezuela at the beginning of 1999, is considered to include the elections of Evo Morales in Bolivia, Rafael Correa in Ecuador, and Daniel Ortega in Nicaragua, all in 2006, and to a lesser extent Luiz Ignacio Lula da Silva in Brazil in 2002, Nestor Kirchner in Argentina in 2003, and Tabar\'{e} Vázquez in Uruguay in 2005.} During 2000--2021, social protection accrued, on average, over 40\% of speech content and remains the dominant policy issue discussed in presidential speeches to this day.		
	
	These patterns reveal how presidents have characterized a country's priorities over time, shifting from military interventions and the development of state capacity, to building physical capital through investments in infrastructure and public services, and finally to promoting human capital through investments in education, health, and social safety nets. Moreover, while idiosyncratic policy issues comprise a non-trivial share of speech content (as reflected in a 5\%--25\% speech content share of the ``Other Topics'' residual category), we also find that a small number of key policy issues make up a significant portion of the priorities expressed by presidents. 
	
	In Appendix \ref{app:empirical}, we conduct a series of robustness checks. First, we vary the number of LDA topics and show that the patterns are similar across the number of topics. Second, we divide our sample into four sub-periods and estimate LDA separately for each period. We find that the list of topics and their evolution over time are remarkably consistent with our baseline results.

	%%%%%%%%%%%%%%%%%%%%%%%%%%%%%%%%%%%%%%%%%%%%%%%%%%%%%%%%%%%%%%%%%%%%%%%%%%%%%%%
	\section{Conclusion} \label{sec:conclusions}
	%%%%%%%%%%%%%%%%%%%%%%%%%%%%%%%%%%%%%%%%%%%%%%%%%%%%%%%%%%%%%%%%%%%%%%%%%%%%%%%	
	
	In this paper, we combine data from presidential speeches with natural-language-processing algorithms to measure the expressed priorities of presidents. We show that high-dimensional speeches can be characterized by a compact set of policy issues. Consistent with political attention theories positing that issue attention remains stable for long periods \citep[e.g.,][]{baumgartner2010agendas}, we find that expressed presidential priorities shift slowly over time and span multiple electoral cycles. Moreover, in line with previous work relating political actors' traits to their policy preferences \citep[e.g.,][]{clots2012female, besley2006health}, we find that presidential traits, such as age and gender, are correlated with issue attention. These novel empirical facts offer insights for political agenda-setting theories and expand our understanding of presidential issue attention.
	
	Future research could leverage the dataset of presidential speeches to measure whether the policy prioritization of an issue in a given country has spillover effects on the likelihood of other countries discussing the same issue in subsequent years \citep{buera2011learning}. In addition, our dataset can be used to learn how the conceptualization of different policy tools (such as personal income taxes or environmental regulations) changes over time and how these correlate with measures of actual policy implementation.

%	\clearpage
	\begin{singlespace}
		\bibliographystyle{chicago}
		\bibliography{cer-speeches}
	\end{singlespace}

	\clearpage
	\begin{center}\noindent {\LARGE \textbf{Appendix --- For Online Publication}}\end{center}
	\appendix
	
	%%%%%%%%%%%%%%%%%%%%%%%%%%%%%%%%%%%%%%%%%%%%%%%%%%%%%%%%%%%%
	\section{Additional Figures and Tables} \label{app:add-figs}
	%%%%%%%%%%%%%%%%%%%%%%%%%%%%%%%%%%%%%%%%%%%%%%%%%%%%%%%%%%%%
	
	\setcounter{table}{0}
	\setcounter{figure}{0}
	\setcounter{equation}{0}	
	\renewcommand{\thetable}{A\arabic{table}}
	\renewcommand{\thefigure}{A\arabic{figure}}
	\renewcommand{\theequation}{A\arabic{equation}}

	\begin{figure}[H]
		\caption{Perplexity and number of topics discussed in presidential speeches}\label{fig:perplexity}
		\centering
		\includegraphics[width=.8\linewidth]{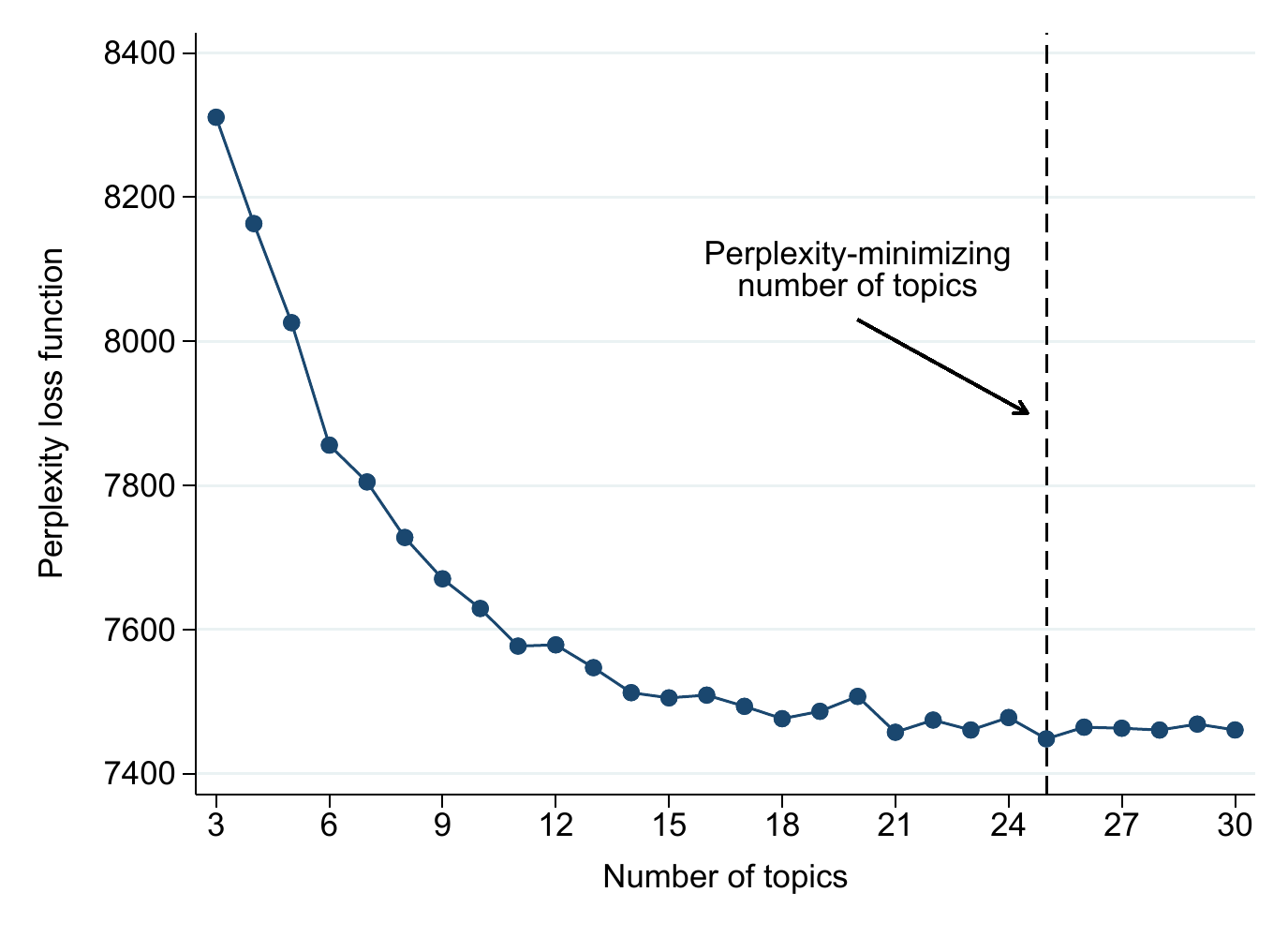}
		\footnotesize
		\singlespacing \justify \textit{Notes:} This figure shows the estimated perplexity loss function against the number of topics in our sample of presidential speeches. To construct this figure, we follow a three-step process: (i) randomly partition our sample into a training set (90\% of speeches) and a hold-out test set (10\% of speeches), (ii) iteratively implement LDA on the training set using a different number of topics in each iteration, and (iii) compute the perplexity of each model. See Appendix \ref{app:perplexity} for details about perplexity. %The vertical dashed line indicates the number of topics that minimizes the perplexity loss function. 
	\end{figure}

	%-------------- Appendix Tables ------------------%
	
	%	\clearpage		
	\begin{table}[H]{\footnotesize
			\begin{center}
				\caption{Exercepts from presidential speeches illustrating the content of each topic}\label{tab:excerpts}
				\begin{tabular}{p{10em}p{4em}p{28em}}
					\midrule
					Speech  & Topic & Excerpt \\ 
					& Prob. &         \\ 
					(1) & (2) & (3) \\ \midrule 
					
					\multicolumn{3}{l}{\hspace{-1em} \textbf{Panel A. War and Patriotism}} \\ \addlinespace		
					
					Peru, 1835 & 0.63 & The triumph of the United Army in Yanacocha, the new order following this battle, the performance of the United Army, the fixing of the national treasury, the peace in the towns and respect towards people and property, are testimony in favor of the government and against its detractors. \\ \addlinespace
					
					Venezuela, 1850 & 0.57 & Tireless and tenacious, the enemies of Venezuela have made formidable efforts to bring down the institutions and introduce a system that is diametrically opposed to the nation's will. [...] The government, always behaving frankly and generously, never withdrew its merciful hand and attempted to attract the strayed to the heart of the community in order to extirpate any germ of new revolts and rebellions, and cement in this way the peace that the towns need so much. \\ \addlinespace \midrule
					
					\multicolumn{3}{l}{\hspace{-1em} \textbf{Panel B. Public Administration}} \\ \addlinespace		
					
					Costa Rica, 1897 & 0.78 & The practice of removing Judges or Mayors only if there is a judicial decision to imprison them is another obstacle, perhaps larger in magnitude, that hinders good public service. [...] this reform would enable more effective and expedited action in the internal order of the Judiciary and would stimulate lower-ranking public servants to strictly comply with the delicate mission they have been entrusted with. \\ \addlinespace
					
					Peru, 1895 & 0.75 & The task of investigating the main administrative branches required staff exclusively dedicated to it, and to satisfy this important requirement, in addition to the inspectors of the Treasury and Customs, three investigative commissions were created: one tasked with examining the fiscal contracts of the previous administration; another with investigating public expenditures; and the third tasked with inspecting the state of the Callao Customs and the cause of its recent weakness. \\  \addlinespace \midrule
					
					\multicolumn{3}{l}{\hspace{-1em} \textbf{Panel C. Infrastructure}} \\ \addlinespace		
					
					Peru, 1959 & 0.53 & During the third year of my administration, 1,379km of roads have been constructed, with an investment of 234 million soles, and work has been done to maintain 12,500km of roads at a cost of more than 60 million soles... These construction projects will allow the promotion and introduction of basic industries, and will open new work centers. \\ \addlinespace
					
					Argentina, 1940 & 0.49 & Concerning the public works, the Executive has proposed to carry out actions oriented towards establishing a uniform type of construction within a model that satisfies the technical requirements of its intended use, procuring the suppression of monumental buildings... \\  \addlinespace \midrule
					
				\end{tabular}%
			\end{center}
		}
	\end{table}

	\begin{table}[H]{\footnotesize	\ContinuedFloat
			\begin{center}
				\caption{Exercepts from presidential speeches illustrating the content of each topic (continued)}
				\begin{tabular}{p{10em}p{4em}p{28em}}
					\midrule
					Speech  & Topic & Excerpt \\ 
					& Prob. &         \\ 
					(1) & (2) & (3) \\ \midrule 
					
					\multicolumn{3}{l}{\hspace{-1em} \textbf{Panel D. Rights and Freedom}} \\ \addlinespace		
					
					Argentina, 1953 & 0.69 & In order to establish political sovereignty, I gave every Argentinian individual freedom in the effective enjoyment of all their rights, which arise from the dignity that can only be enjoyed by men who have been economically liberated by social justice... \\ \addlinespace	
					
					Peru, 1973 & 0.62 & This means recognizing the right of others to think and act differently from us, and consequently, to organize themselves politically with complete freedom within a plurality of alternatives. Our Revolution represents one of these alternatives. \\ \addlinespace \midrule
					
					\multicolumn{3}{l}{\hspace{-1em} \textbf{Panel E. Economic Development}} \\ \addlinespace		
					
					Mexico, 1983 & 0.70 & We have presented and initiated actions to induce qualitative changes to the economic structure, to revise attitudes and update styles in order to improve the orientation and quality of development, and to transform it into a steady and sustained process \\ \addlinespace	
					
					Venezuela, 1989 & 0.58 & We are currently on the path towards economic growth and social progress, so that the decade of the 90s becomes the decade of development. \\ \addlinespace \midrule
					
					\multicolumn{3}{l}{\hspace{-1em} \textbf{Panel F. Social Protection}} \\ \addlinespace		
					
					Mexico, 2008 & 0.72 & I have no doubt that we will continue to work intensely to build a fairer future for you and your family; to build a more humane Mexico, a Mexico with sustainable human development; a Mexico without extreme poverty; a Mexico with health and education for all. \\ \addlinespace

					Chile, 2012 & 0.61 & Education is the key engine of development and social mobility. It is the mechanism needed for the talent and merit to emerge. It is the great instrument for the construction of a country of opportunities. For this reason, the battle for development and against poverty will be won or lost in the classrooms. \\ \addlinespace \midrule
					
				\end{tabular}%
			\end{center}
			\begin{singlespace} \vspace{-.5cm}
				\noindent \justify \textit{Notes:} This table shows speech excerpts that illustrate the six main topics identified by LDA. We show excerpts from speeches with a high topic probability. We estimate the topics and their probability distributions using a Latent Dirichlet Allocation (LDA) algorithm (see Section \ref{sub:topic-models} and Appendix \ref{app:lda}). To choose the number of topics, we follow a criterion of perplexity minimization (see Appendix \ref{app:perplexity}). We present only topics whose probability exceeds 20\% in at least one decade. Topics are defined by their top occurring keywords (see Table \ref{tab:topics-speeches} for the top ten keywords that define the topics in the figure). We manually labeled topics based on the top keywords. 
			\end{singlespace}
			
		}
	\end{table}

	%\begin{landscape}
	\begin{sidewaystable}
		\begin{table}[H]{\footnotesize			
				\begin{center}
					\caption{Robustness of the president-level correlates of expressed priorities to alternative specifications} \label{tab:det-issue-rob}
					\newcommand\w{2.3}
					\begin{tabular}{l@{}lR{\w cm}@{}L{0.45cm}R{\w cm}@{}L{0.45cm}R{\w cm}@{}L{0.45cm}R{\w cm}@{}L{0.45cm}R{\w cm}@{}L{0.45cm}R{\w cm}@{}L{0.45cm}R{\w cm}@{}L{0.45cm}}
						\midrule
						&& \multicolumn{12}{c}{Dependent variable: Fraction of a speech discussing...} \\ \cmidrule{2-14}
						&& War and    && Public          && Infrastructure    && Rights and  && Economic && Social   \\					
						&& Patriotism && administration  &&  && freedom     && development   && protection \\				
						&& (1)        && (2)             && (3)       && (4)         &&   (5)  &&   (6)      \\
						\midrule
						
						\input{results/issue-det}\midrule
						\input{results/issue-det-n}\midrule 		
					\end{tabular}
				\end{center}
				\begin{singlespace}  \vspace{-.7cm}
					\noindent \justify \textit{Notes:} This table is analogous to Table \ref{tab:det-issue}, but all variables enter simultaneously in the regression equation. Heteroskedasticity-robust standard errors in parentheses. $^{***}$, $^{**}$, and $^{*}$ denote significance at the 1\%, 5\% and 10\% levels, respectively. 
				\end{singlespace} 	
			}
		\end{table}
	\end{sidewaystable}
	%\end{landscape}	

	\clearpage
	%%%%%%%%%%%%%%%%%%%%%%%%%%%%%%%%%%%%%%%%%%%%%%%%%%%%%%%%%%%%
	\section{Data Appendix} \label{app:data}
	%%%%%%%%%%%%%%%%%%%%%%%%%%%%%%%%%%%%%%%%%%%%%%%%%%%%%%%%%%%%
	
	\setcounter{table}{0}
	\setcounter{figure}{0}
	\setcounter{equation}{0}	
	\renewcommand{\thetable}{B\arabic{table}}
	\renewcommand{\thefigure}{B\arabic{figure}}
	\renewcommand{\theequation}{B\arabic{equation}}	
	
	\subsection{Presidential Speeches Inclusion Criteria} \label{app:data-speeches}
	
	We included countries into our sample based on two criteria. First, we restricted our search to Spanish-speaking countries in order to be able to pool speeches for text analysis. Second, we focused only on countries that had an annual, constitutionally-mandated presidential speech in which the president gives an overview of the work the government has performed in each legislative session, as well as an outline of the policy goals and priorities for the future. Appendix Table \ref{tab:app-countries} lists the Spanish-speaking Latin American countries that have this constitutional mandate, as well as the specific articles in the constitution that establish this requirement. Our sample is composed of the subset of countries for which we could locate speeches across at least two decades.

	\begin{table}[H]{\footnotesize		
			\begin{center}
				\caption{\centering Countries with constitutionally-mandated presidential speeches} \label{tab:app-countries}
				\begin{tabular}{lc}
					\hline
					{Country} & {Article} \\
					(1) & (2) \\
					\hline
					Argentina & 99, subsection 8 \\
					Bolivia & 96, subsection 10 \\
					Colombia & 189, subsection 12 \\
					Costa Rica & 139, subsection 4 \\
					Chile & 24 \\
					Ecuador & 171, subsection 7 \\
					Guatemala & 183, subsection i \\
					Honduras & 183, subsection i \\
					Mexico & 69 \\
					Nicaragua & 150, subsection 15 \\
					Panama & 178, subsection 5 \\
					Paraguay & 238, subsection 8 \\
					Peru & 118, subsection 7 \\
					Venezuela & 237 \\
					Dominican Republic & 55, subsection 22 \\
					Uruguay & 168, subsection 5 \\
					\hline
				\end{tabular}
			\end{center}
			\begin{singlespace} \vspace{-.5cm}
				\noindent \justify \textit{Notes:} This table lists the Spanish-speaking Latin American countries with a constitutionally-mandated state-of-the-union-style speech and the article in the constitution that establishes this requirement.
			\end{singlespace}
		}
	\end{table}

	\subsection{About the Missing Speeches}
	
\begin{figure}[H]
		\caption{Cumulative number of presidential speeches by country over time}\label{fig:speeches}
		\centering
		\includegraphics[width=.8\linewidth]{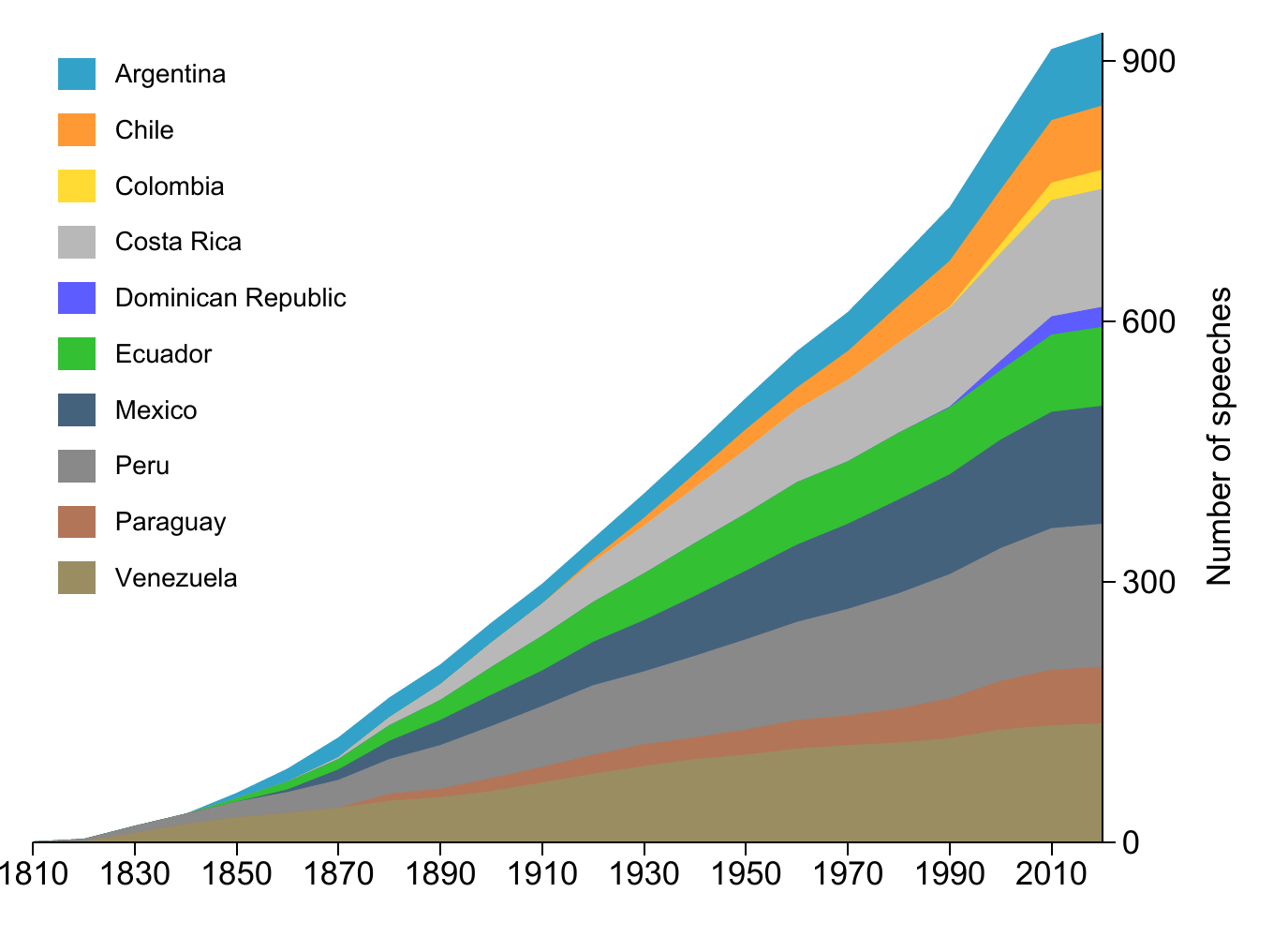}
		\footnotesize \vspace{-.1cm}	
		\singlespacing \justify \textit{Notes:} This figure shows the cumulative number of speeches in the countries in our sample. See Appendix \ref{app:data} for details on the data.
	\end{figure}

	Appendix Figure \ref{fig:speeches} shows the cumulative sum of speeches by year. 
	
	Although we made efforts to compile as many presidential speeches as possible, there are a number of countries for which the speeches from certain years are missing. In some years, presidential speeches are missing because no speech was delivered. For example, in the case of Paraguay, the president did not deliver a speech from 1940--1948. Political turmoil, such as a coup or an ongoing revolution, is another probable cause, particularly for missing speeches from the 1970s. Given the multiple coups and revolts in the region during this time, some of the missing speeches are likely associated with this turbulent political context.
	
	It is hard to establish a cause with certainty for other missing speeches. In some cases, the congressional libraries could not locate the physical copies; in other cases, the quality of the original manuscripts was too low for proper digitalization. At times, particularly in the case of Venezuela, we could not find publications containing the compilations of the presidents' speeches for specific years.

\clearpage
%%%%%%%%%%%%%%%%%%%%%%%%%%%%%%%%%%%%%%%%%%%%%%%%%%%%%%%%%%%%
\section{Empirical Appendix} \label{app:empirical}
%%%%%%%%%%%%%%%%%%%%%%%%%%%%%%%%%%%%%%%%%%%%%%%%%%%%%%%%%%%%

\setcounter{table}{0}
\setcounter{figure}{0}
\setcounter{equation}{0}	
\renewcommand{\thetable}{C\arabic{table}}
\renewcommand{\thefigure}{C\arabic{figure}}
\renewcommand{\theequation}{C\arabic{equation}}

%	\subsection{Robustness of Expressed Priorities Over Time} \label{sub:cycle-robustness}

\begin{figure}[H]
	\caption{Robustness of topics discovered by LDA to the number of topics in the speeches}\label{fig:topics-speeches-t}
	\centering
	\begin{subfigure}[t]{.65\textwidth}
		\caption*{Panel A. Number of topics = 5}\label{fig:topics-speeches-t5}
		\centering
		\includegraphics[width=\linewidth]{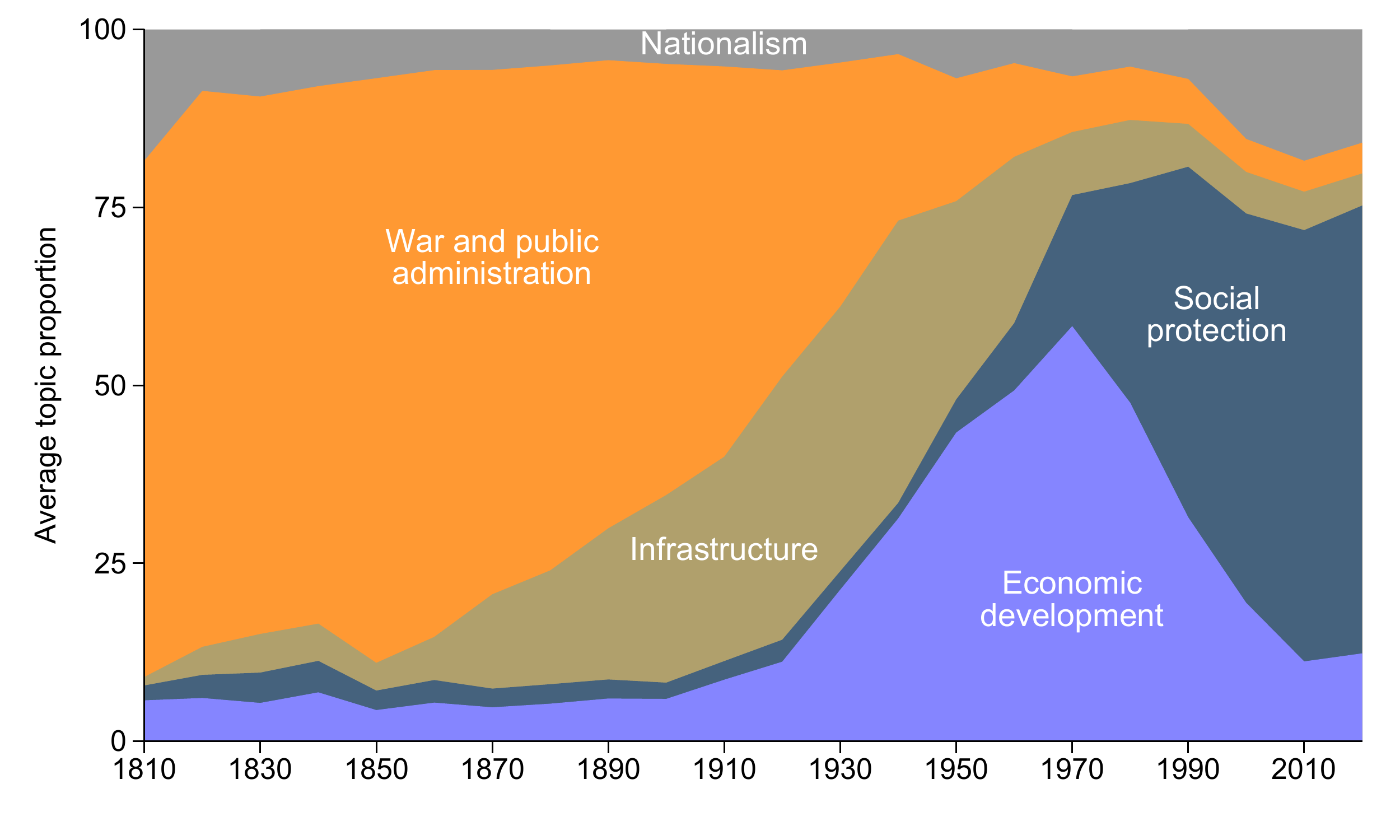}
	\end{subfigure}
	\hfill		
	\begin{subfigure}[t]{0.65\textwidth}
		\caption*{Panel B. Number of topics = 15}\label{fig:topics-speeches-t15}
		\centering
		\includegraphics[width=\linewidth]{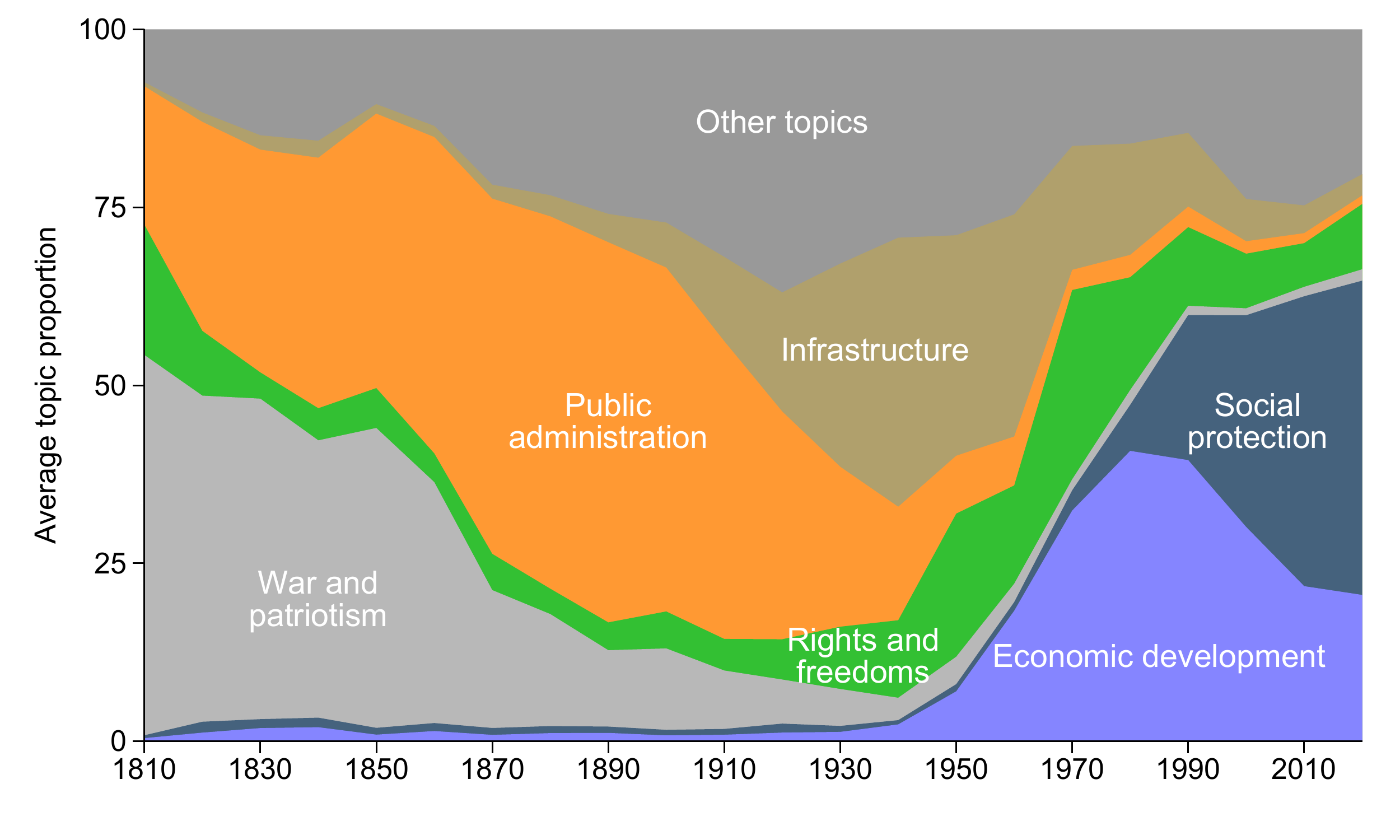}
	\end{subfigure}
	\hfill				
	\begin{subfigure}[t]{0.65\textwidth}
		\caption*{Panel C. Number of topics = 45}\label{fig:topics-speeches-t45}
		\centering
		\includegraphics[width=\linewidth]{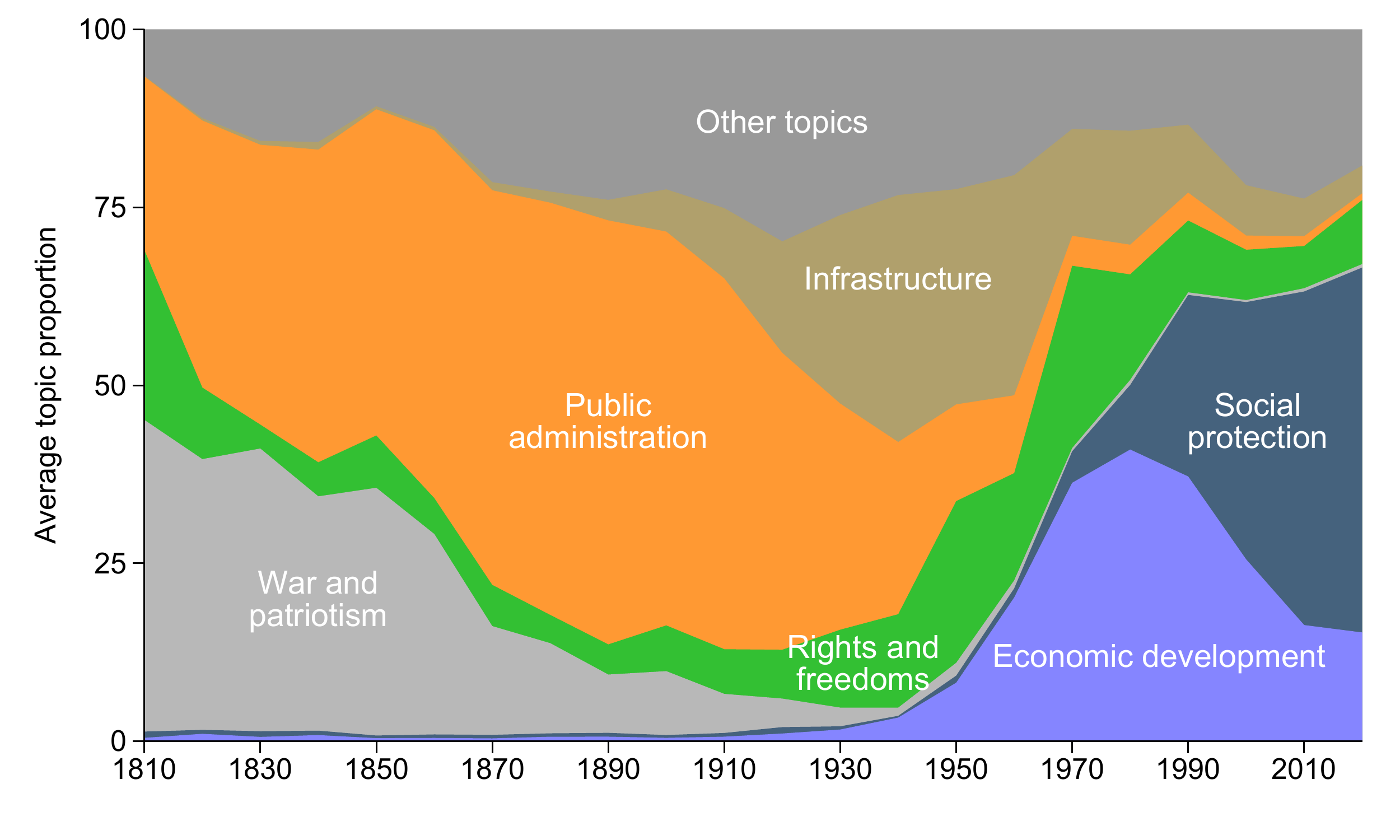}
	\end{subfigure}
	{\footnotesize
		\singlespacing \justify
		
		\textit{Notes:} This figure shows the distribution of topics across decades. We estimate the topics and their probability distributions using a Latent Dirichlet Allocation (LDA) algorithm (see Section \ref{sub:topic-models} and Appendix \ref{app:lda}). In Panel A, we set the number of topics equal to 5; in Panel B, equal to 15; in Panel C, equal to 45. We present only topics whose probability exceeds 20\% in at least one decade. The rest of the topics are grouped together in the category labeled as ``Other topics.'' Topics are defined by their top occurring keywords (see Table  \ref{tab:topics-speeches} for the top ten keywords that define the topics in the figure). We manually labeled topics based on the top keywords. To construct the figure, we pool speeches at the decade level and compute the average topic probabilities in each decade. 
		
	}
\end{figure}

We assess the robustness of the Section \ref{sub:issue-cycle} results in two ways. First, we vary the number of topics to be discovered by the LDA algorithm. We repeat our analysis using 5, 15, and 45 topics (Appendix Table \ref{tab:topics-speeches-t} and Appendix Figure \ref{fig:topics-speeches-t}). The main topics uncovered by LDA are very similar when varying the number of topics and the evolution of these topics over time mirrors the one in the baseline analysis.

The second robustness check relates to a shortcoming of how LDA operates with time-series data. LDA assumes that the vocabulary is fixed over time \citep{blei2006dynamic}. However, a significant change in language could affect topic discovery and assignment. To deal with this, we follow  \cite{kim2011topic} and partition our sample into different periods. Then, we estimate the topics separately for each period. This allows LDA to discover the topics that dominated each period using only the words observed in that period instead of the words found in the entire vocabulary in our sample.

\begin{figure}[H]
	\caption{Robustness of topics discovered to conducting LDA by subperiods}\label{fig:topics-periods}
	\centering
	\includegraphics[width=.75\linewidth]{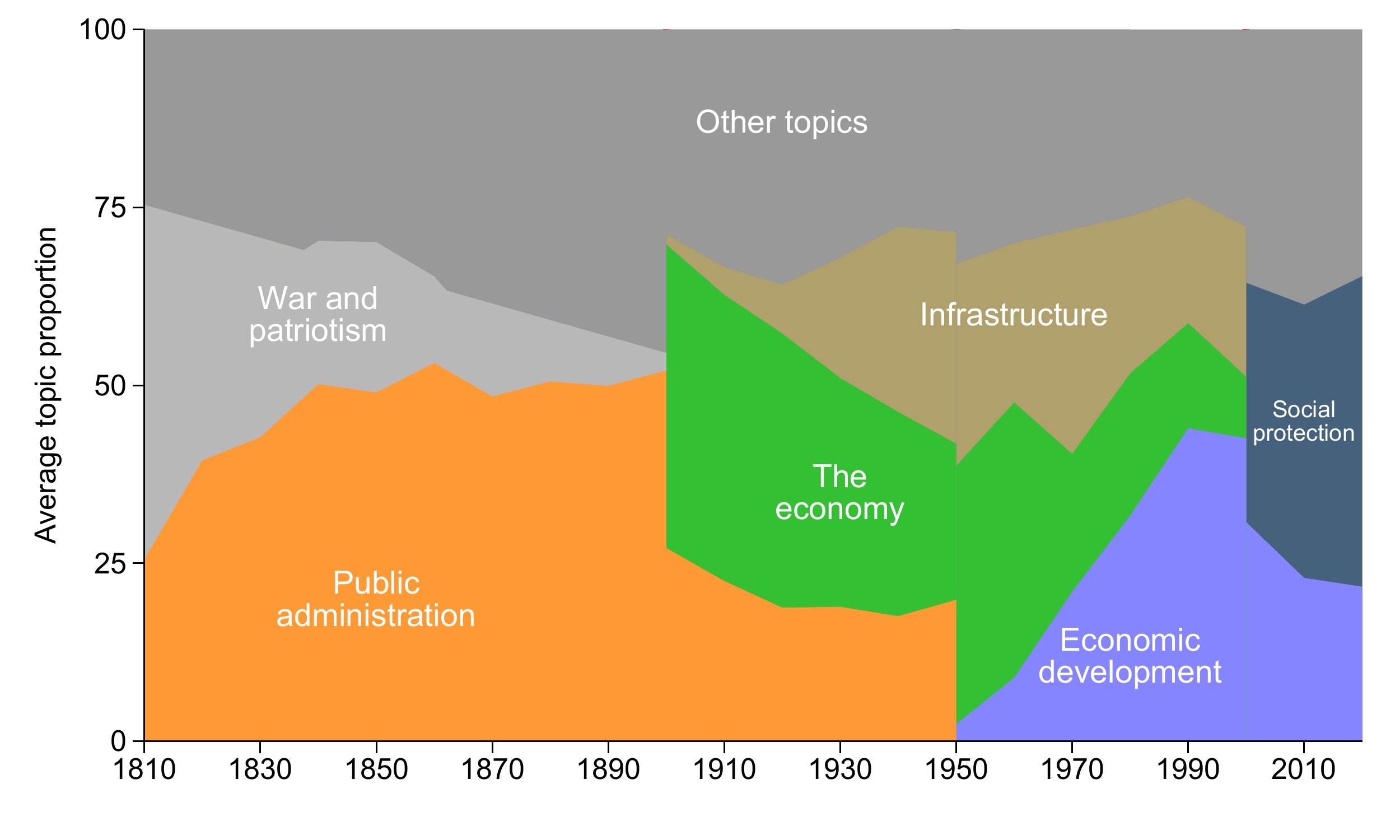}	
	\footnotesize
	\singlespacing \justify \textit{Notes:} This figure shows the distribution of topics calculated separately in four periods (with a similar number of observations per period): 1819--1900, 1901--1950, 1951--2000, and 2001--2021. We estimate the topics and their probability distributions using a Latent Dirichlet Allocation (LDA) algorithm (see Section \ref{sub:topic-models} and Appendix \ref{app:lda}). We fix the number of topics to 25 across all periods (the number of topics that minimizes perplexity for the pooled sample). We present only topics whose probability exceeds 20\% in at least one decade. The rest of the topics are grouped together in the category labeled as ``Other topics.'' Topics are defined by their top occurring keywords (see Table  \ref{tab:topics-periods} for the top ten keywords that define the topics in the figure). We manually labeled topics based on the top keywords. To construct the figure, we pool speeches at the decade level and compute the average topic probabilities in each decade. 
	
\end{figure}

We divide our sample into four periods (with a similar number of observations per period): 1819--1900, 1901--1950, 1951--2000, and 2001--2021, and estimate the LDA algorithm on the speeches of each subperiod. Appendix Table \ref{tab:topics-periods} shows the main topics in each period and Appendix Figure \ref{fig:topics-periods} juxtaposes the evolution of these topics across periods. The list of topics is very similar to the list LDA yields when pooling all the periods together.

\begin{table}[H]\footnotesize
	\protect\caption{Top ten keywords defining main topics}\label{tab:topics-speeches-t}
	
	\begin{centering} 
		\textbf{Panel A. LDA using 5 topics}
		\begin{tabular}{lcccccc}
			\addlinespace \midrule
			& \multicolumn{5}{c}{Topic name} \\\cmidrule{2-6}				
			& War and Public &                &             & Social         & Economic \\					
			& administration & Infrastructure & Nationalism & protection     & development  \\									
			& (1) & (2) & (3) & (4) & (5)  \\
			\midrule 	
			\input{results/topics-t5.tex} \midrule  \addlinespace \addlinespace
		\end{tabular}
		
		\textbf{Panel B. LDA using 15 topics}
		\begin{tabular}{lccccccc}
			\addlinespace \addlinespace \midrule
			& \multicolumn{6}{c}{Topic name} \\\cmidrule{2-7}				
			& War and    & Public         &                & Rights and  & Economic    & Social \\					
			& Patriotism & administration & Infrastructure & freedom     & development & protection \\									
			& (1) & (2) & (3) & (4) & (5) & (6) \\
			\midrule 	
			\input{results/topics-t15.tex} \midrule \addlinespace \addlinespace	 
		\end{tabular}

		\textbf{Panel C. LDA using 45 topics}
		\begin{tabular}{lccccccc}
			\addlinespace \addlinespace \midrule
			& \multicolumn{6}{c}{Topic name} \\\cmidrule{2-7}				
			& War and    & Public         &                & Rights and  & Economic    & Social \\					
			& Patriotism & administration & Infrastructure & freedom     & development & protection \\									
			& (1) & (2) & (3) & (4) & (5) & (6) \\
			\midrule 	
			\input{results/topics-t45.tex} \midrule \addlinespace
		\end{tabular}
		\par\end{centering} \vspace{-.3cm}
	
	\singlespacing\justify
	
	\textit{Notes}: This table lists the top ten keywords that define the topics in Figure \ref{fig:topics-speeches-t}. We only list topics whose probability exceeds 20\% in at least one decade. Each column header shows the manually-assigned label of each topic. We recover the topics and their probability distributions using a Latent Dirichlet Allocation (LDA) algorithm (see Section \ref{sub:topic-models} and Appendix \ref{app:lda}).
	
\end{table}

%	\begin{landscape}
	\begin{sidewaystable}
		\begin{table}[H]{\scriptsize
				\begin{centering} 
					\protect\caption{Top keywords defining main  topics} \label{tab:topics-periods}
					\newcommand\w{1.5}
					\begin{tabular}{lcccccccccccccc}
						\addlinespace \addlinespace \midrule
						& \multicolumn{2}{c}{1819--1900} && \multicolumn{3}{c}{1901--1950} &&  \multicolumn{3}{c}{1951--2000} && \multicolumn{2}{c}{2001--2021} \\\cmidrule{2-3} \cmidrule{5-7} \cmidrule{9-11}\cmidrule{13-14}				
						& War and    & Public         &&                &             & Public          && Economic      &              &               && Economic   & Social \\					
						& Patriotism & administration && Infrastructure & The economy & administration  && development & The economy & Infrastructure && development & protection \\									
						& (1) & (2) & & (3) & (4) & (5) && (6) & (7) & (8) && (9) & (10) \\
						\midrule 	
						\input{results/topics-periods.tex} \midrule
						
					\end{tabular}
					\par\end{centering}
				
				\singlespacing\justify \footnotesize
				\textit{Notes}: This table lists the top ten keywords that define the topics in each in Appendix Figure \ref{fig:topics-periods}. We only list topics whose probability exceeds 20\% in at least one decade. Each column header shows the manually-assigned label of each topic. We recover the topics and their probability distributions using a Latent Dirichlet Allocation (LDA) algorithm (see Section \ref{sub:topic-models} and Appendix \ref{app:lda}).
				
			}
		\end{table}
	\end{sidewaystable}
	%	\end{landscape}	

\clearpage

\clearpage
%%%%%%%%%%%%%%%%%%%%%%%%%%%%%%%%%%%%%%%%%%%%%%%%%%%%%%%%%%%%
\section{Latent Dirichlet Allocation Algorithm} \label{app:lda}
%%%%%%%%%%%%%%%%%%%%%%%%%%%%%%%%%%%%%%%%%%%%%%%%%%%%%%%%%%%%

\setcounter{table}{0}
\setcounter{figure}{0}
\setcounter{equation}{0}	
\renewcommand{\thetable}{D\arabic{table}}
\renewcommand{\thefigure}{D\arabic{figure}}
\renewcommand{\theequation}{D\arabic{equation}}	

\subsection{Description of the LDA Algorithm} \label{app:lda-summ}

Topic models are statistical algorithms that uncover the main topics contained in a collection of documents \citep{blei2012probabilistic}. The most widely used topic model is the Latent Dirichlet Allocation (LDA) algorithm \citep{blei2003latent}.	The fundamental assumption of the LDA algorithm is that the observed documents were generated through a particular probabilistic generative process.\footnote{Another important assumption is the ``exchangeability'' or ``bag-of-words'' assumption, which means that the order in which the words appear in the document is not important; LDA relies on term frequencies instead.} However, the parameters or ``recipe'' of this generative process are hidden or ``latent.'' This defines the key inferential task of LDA: estimating the ``latent'' structure—the topics and the topic composition of each document. LDA performs this task by working through the generative process in reverse. That is, it uses the observed words in each document to estimate the parameters of the generative process that are most likely to have generated the observed collection of documents.

The generative process can be described as follows:

\begin{enumerate}
	\item For each topic, decide what words are likely.
	\item For each document:
	\begin{enumerate}
		\item Decide what proportions of topics should be in the document, say 20\% topic A, 50\% topic B, 30\% topic C.
		\item For each word:
		\begin{enumerate}
			\item Choose a topic. Based on the topic proportions from step 2.a., topic A is more likely to be chosen.
			\item Given this topic, choose a likely word (generated in step 1).
		\end{enumerate}
	\end{enumerate}
\end{enumerate}

We can describe this process more formally using the model parameters and corresponding probability distributions. After specifying a number of topics $k$:

\begin{enumerate}
	\item For each topic $k$, draw a distribution over words $\varphi_k$ according to a Dirichlet distribution $\sim$ Dir($\beta$), where $\beta$ is the parameter of the Dirichlet prior on the per-topic word distribution.\footnote{The beta parameter represents the ``prior'' belief about the per-topic word distributions. A high beta value means that each topic is likely to be made up of most of the words in the corpus, whereas a low beta means that each topic will have fewer words.}
	\item For each document $d$:
	\begin{enumerate}
		\item Draw a vector of topic proportions $\theta_d$ according to a Dirichlet distribution $\sim$ Dir($\alpha$), where $\alpha$ is the parameter of the Dirichlet prior on the per-document topic distribution.\footnote{The alpha parameter represents the prior belief about the per-document topic distributions. A high alpha-value means that each document is likely to contain a mixture of most of the topics, and not any single topic specifically, whereas a low alpha-value means that each document is likely to contain fewer topics.}
		\item For each of the $N$ words $w_n$:
		\begin{enumerate}
			\item Draw a topic assignment $z_n$ according to a multinomial distribution $\sim$ Multinomial(
			$\theta$) according to the topic proportion $\theta_d$.
			\item Choose a word $w_n$ from $\Pr(w_n|z_n, \varphi)$, a multinomial probability conditioned on the topic $z_n$.
		\end{enumerate}
		
	\end{enumerate}
	
\end{enumerate}

The key inferential task of LDA consists in performing this assumed generative process in reverse. That is, using the observed documents and words, LDA works backwards to infer the ``latent structure''---the distribution of the parameters $\theta$, $z$, and $\varphi$---that are most likely to have generated the documents in the sample. Where $z$ represents the per-word topic assignments and $\theta$ gives the topic distribution of each document, which indicates the extent to which each document belongs to each topic; $\varphi$ gives the distribution of words in topic $k$, which is used to define the semantic content of each topic. The objective of LDA consists in computing the posterior distribution of these hidden variables given a document and the Dirichlet priors:
\begin{align*}
	\Pr\left(\theta, z, \varphi\middle|w, \alpha, \beta \right) = \frac{\Pr(\theta,z, \varphi| \alpha, \beta)}{\Pr(w | \alpha,\beta)}.
\end{align*}

Estimating the maximum likelihood of the model and the distributions of the hidden variables requires marginalizing over the hidden variables to obtain the model's probability for a given corpus $w$ and priors $\beta$ and $\alpha$.
\begin{align*}
	\Pr\left(w\middle| \alpha, \beta \right) = \frac{\Gamma(\sum_i \alpha_i)}{\prod_i \Gamma(\alpha_i)} \ \int\left(\prod_{i=1}^{k}\theta_i^{\alpha_i-1}\right)\left(\prod_{n=1}^{N}\sum_{i=1}^{k}\prod_{j=1}^{V}\left(\theta_i\beta_{ij} \right)^{w_n^j} \right) d\theta.
\end{align*}

These distributions are intractable to compute, requiring the use of other approximate inference algorithms. Although in the first introduction of LDA, Blei et.al. (2003) relied on a Variational Bayes approximation of the posterior distribution, we use Collapsed Gibbs sampling as our inference technique---a commonly used alternative introduced by Griffith and Steyvers (2004). 

\subsection{Collapsed Gibbs Sampling}

The Collapsed Gibbs sampling algorithm is a common Markov Chain Monte Carlo (MCMC) algorithm that is used to approximate posterior distributions when these cannot be directly computed. The idea is to iteratively generate posterior samples by looping through each variable to sample from its conditional distribution while retaining the values of all other variables fixed in each iteration \citep{yildirim2012bayesian}. Essentially, we simulate posterior samples by sweeping through all the posterior conditionals, one random variable at a time. Because we initiate the algorithm with random values, the samples simulated at the early iterations are likely not close to the true posteriors. However, the process eventually ``converges'' at the point where the distribution of the samples closely approximates the distribution of the true posteriors. 

In LDA, the variables we want to approximate are the ``latent'' variables $\theta$ and $\varphi$. This is achieved by generating a sequence of samples of topic assignments $z$ for each word $w$. As mentioned above, for each iteration, Gibbs Sampling requires retaining the values of all variables except for one fixed (see \citealp{griffiths2004finding}). Therefore, because words are the only observed variables in LDA, at each iteration, the topic assignment of only one word is updated, while the topic assignments for all other words are assumed to be correct (i.e., remain unchanged). Samples from the posterior distribution $\Pr(z|w)$ are obtained by sampling from: 
\begin{align*}
	\Pr\left(z_i = K\middle|w, z_{-i}\right) \frac{{n}_{-i,K}^{(j)}+ \beta n_{-i,K}^{(di)}+\alpha}{n_{-i,K}^{(.)} + V\beta n_{(-i,.)}^{(di)} + k\alpha},
\end{align*}
where $z_{-i}$ is the vector of current topic assignments of all words except the $i$th word $w_{i}$. The index $j$ indicates that $w_{i}$ is the $j$th term in the entire vocabulary of words in the corpus ($V$); ${n}_{-i,K}^{(j)}$ indicates how often the $j$th term of the vocabulary is currently assigned to topic $K$ without the $i$th word. The dot ``.'' indicates that summation over this index is performed; $d_{i}$ denotes the document in the corpus to which word $w_{i}$ belongs; $\beta$ and $\alpha$ are the hyperparameters of the prior distribution explained above \citep{grun2011topicmodels}.

Intuitively, the algorithm beings by going through each document and randomly assigns each word in the document to one of the K topics. Because these assignments are random, however, they are poor and must be improved on. To improve these topic assignments, for each word $i$ in document $d$, (for each $w_{d,i}$) and for each topic $k$, two values are computed: 1) $\Pr$(topic $k$ $|$ document $d$), or the proportion of words in document $d$ that are currently assigned to topic $k$, and 2) $\Pr$(word $w$ $|$ topic $k$), or the proportion of assignments to topic $k$ over all documents that come from this word $w$. Then, these two proportions are multiplied to get $\Pr$(topic $t$ $|$ document $d$) $\times$ $\Pr$(word $w$ $|$ topic $t$), which in the context of LDA's generative process, gives the probability that topic $k$ generated word $w$. Finally, word $w$ is reassigned to a new topic based on this probability. To put it simply, for each word, its topic assignment is updated based on two criteria: 1) How prevalent is that word across topics? 2) How prevalent are topics in the document? 

As in any Gibbs Sampling algorithm, the above steps are repeated a large number of times. After a large number of iterations, the algorithm converges to a steady state where the topic assignments of each word are close approximations of the true values. At this point, we can finally use these topic assignments to estimate the ``latent'' variables---the posterior of $\theta$ and $\varphi$---given the observed words $w$ and topic assignments $z$:
\begin{align*}
	{\hat{\theta}}_K^d=\ \frac{n_K^{(d)}+\alpha}{n^{(d)}+k\alpha} \quad;\quad {\hat{\varphi}}_K^j=\ \frac{n_K^{(j)}+\beta}{n_K^{(.)}+V\beta} \qquad \text{ for } j = 1, ..., V \text{ and } d = 1, ... D
\end{align*}

With $\theta$ and $\varphi$ estimated, the objective of LDA---extracting topic representations of each document---is achieved.

\subsection{Empirical Implementation}

As a first step before running LDA, to improve the discovery of topics, we follow the standard practice of cleaning our collection of documents. Specifically, we remove all punctuation and numbers, as well as ``stop words”—terms such as articles, conjunctions, and pronouns that are semantically meaningless for defining a topic. Because we are interested in discovering topics that are common across countries, we remove country-specific terms such as ``Peruvians,'' ``Peru,'' or ``Lima,'' which could otherwise bias the topics towards country-specific rather than subject-specific topics.

We rely on Collapsed Gibbs Sampling for the iterative process of topic inference. This approach requires the specification of values for the parameters of the prior distributions---$\beta$ for the per-topic term distributions and $\alpha$ for the per-document topic distributions. Following \cite{griffiths2004finding}, we select the commonly used values of $\alpha = 50/t$ (where $t$ is the number of topics) and $\beta = 0.1$. 

\subsection{Number of Topics to be Discovered} \label{app:perplexity}

One crucial parameter that must be specified is the number of topics to be discovered. To determine the optimal number of topics we rely on a measure known as \textit{perplexity} that is often used in information theory and natural language processing to evaluate how well a model can predict the data, with lower perplexity indicating a better model \citep{blei2003latent}.  Formally, for a test set of $M$ documents, perplexity is defined as:
\begin{align*}
	\text{Perplexity}\left(D_{test}\right)=\exp{\left\{-\frac{\sum_{d=1}^{M}{\log{\Pr(W_d})}}{\sum_{d=1}^{M}N_d}\right\}},
\end{align*}
where $W_d$ represents the words in document $d$ and $N_d$ the number of words. The lower the perplexity, the better the model can predict the data.

To minimize the perplexity, we follow a three-step process: (i) Randomly partition our sample into a training set (90\% of speeches) and a held-out test set (10\% of speeches); (ii) Iteratively implement LDA on the training set using a different number of topics $n \in\{1, \ldots, 30\}$ in each iteration; and (iii) Compute the perplexity of each model, which amounts to evaluating how ``perplexed'' or surprised each ``trained'' model is when presented with the previously unseen test set.

\end{document}

%% file: results/summ-wor.tex
Delivered during 1819--1900 &0.225 &0.418 &933 \\
Delivered during 1901--1950 &0.268 &0.443 &933 \\
Delivered during 1951--2000 &0.301 &0.459 &933 \\
Delivered during 2001--2021 &0.206 &0.404 &933 \\
Words in speech &5,637.7 &5,323.6 &933 \\

%% file: results/summ-lea.tex
Age &54.8 &9.6 &780 \\
Male &0.976 &0.154 &780 \\
Days in office &2,776.2 &2,925.1 &780 \\
Democratically-elected &0.516 &0.500 &902 \\
President's party controls both Houses &0.458 &0.499 &306 \\

%% file: results/topics-allspeeches.tex
1 &Peace &Public &National &Town &Developing &Health \\
2 &Towns &National &Plays &Politics &National &Family \\
3 &Homeland &Service &Production &Social &Social &Education \\
4 &War &Plays &Construction &Life &Program &Program \\
5 &Army &Management &Education &Right &Politics &Social \\
6 &National &Executive &Services &Liberty &Sector &Quality \\
7 &Nation &Right &Social &National &System &Safety \\
8 &Liberty &Interests &Plan &Homeland &Means &Poverty \\
9 &Town &Public &Service &Nation &Process &Right \\
10 &Law &Order &Activities &Economical &Increase &Investment \\

%% file: results/issue-det-democracy.tex
Democracy&&$-$0.093&\sym{***}&$-$0.171&\sym{***}&0.012&&0.013&&0.122&\sym{***}&0.167&\sym{***}\\
&&(0.008&)&(0.014&)&(0.009&)&(0.008&)&(0.010&)&(0.010&)\\

%% file: results/issue-det-age.tex
Age&&$-$0.001&\sym{***}&$-$0.005&\sym{***}&0.001&&0.000&&0.003&\sym{***}&0.002&\sym{***}\\
&&(0.000&)&(0.001&)&(0.000&)&(0.000&)&(0.001&)&(0.001&)\\

%% file: results/issue-det-female.tex
Female&&$-$0.034&\sym{***}&$-$0.215&\sym{***}&$-$0.099&\sym{***}&$-$0.016&&0.060&\sym{***}&0.251&\sym{***}\\
&&(0.004&)&(0.009&)&(0.009&)&(0.038&)&(0.023&)&(0.046&)\\

%% file: results/issue-det-n.tex
\(N\) (speeches)&&780&&780&&780&&780&&780&&780&\\
Mean DV&&0.039&&0.228&&0.143&&0.106&&0.147&&0.096&\\

%% file: results/issue-det.tex
Democracy&&$-$0.040&\sym{***}&$-$0.125&\sym{***}&$-$0.003&&0.007&&0.109&\sym{***}&0.132&\sym{***}\\
&&(0.005&)&(0.015&)&(0.010&)&(0.009&)&(0.012&)&(0.009&)\\
Age&&$-$0.001&\sym{***}&$-$0.004&\sym{***}&0.001&&0.000&&0.002&\sym{***}&0.001&\sym{***}\\
&&(0.000&)&(0.001&)&(0.000&)&(0.000&)&(0.001&)&(0.001&)\\
Female&&$-$0.015&\sym{***}&$-$0.155&\sym{***}&$-$0.098&\sym{***}&$-$0.019&&0.009&&0.190&\sym{***}\\
&&(0.003&)&(0.011&)&(0.010&)&(0.039&)&(0.024&)&(0.046&)\\

%% file: results/topics-t5.tex
1 &Executive &Plays &Village &Program &National \\
2 &Public &National &World &Developing &Social \\
3 &Nation &Service &National &Social &Politics \\
4 &National &Construction &Homeland &National &Developing \\
5 &Right &Services &Social &Health &Production \\
6 &Administration &Department &Revolution &Sector &Village \\
7 &Plays &Works &History &Education &Economical \\
8 &Peace &School &Plan &System &Life \\
9 &Order &Schools &Life &Increase &Program \\
10 &Towns &Federal &Countries &Investment &Means \\

%% file: results/topics-t15.tex
1 &Peace &Public &National &Village &Developing &Health \\
2 &Towns &Administration &Plays &Politics &National &Family \\
3 &Homeland &Service &Production &Social &Social &Education \\
4 &National &Plays &Social &Life &Program &Quality \\
5 &War &National &Services &Liberty &Sector &Poverty \\
6 &Nation &Executive &Construction &Right &Politics &Draft \\
7 &Army &Right &Activities &National &System &Security \\
8 &Law &Interests &Education &Economical &Means &Right \\
9 &Liberty &Order &Plan &Homeland &Process &Investment \\
10 &Patriotism &Instruction &Service &Democracy &Increase &Woman \\

%% file: results/topics-t45.tex
1 &Peace &Public &National &Village &Developing &Health \\
2 &Homeland &Plays &Plays &Politics &National &Education \\
3 &Towns &National &Construction &Social &Social &Family \\
4 &Army &Service &Production &Life &Program &Program \\
5 &War &Administration &Services &Right &Politics &Investment \\
6 &Nation &Executive &Education &Liberty &Sector &Quality \\
7 &National &Right &Social &National &System &Security \\
8 &Liberty &Public &Service &Homeland &Means &Right \\
9 &Province &Interests &Plan &Nation &Process &Poverty \\
10 &Honor &Relations &Ministry &Economical &Increase &Social \\

%% file: results/topics-periods.tex
1 &Towns &National & &Service &National &Right & &Developing &Village &National & &National &Family \\
2 &Homeland &Administration & &Plays &Social &Village & &Program &Social &Plays & &Social &Right \\
3 &Liberty &Public & &National &Production &Life & &National &Politics &Construction & &Program &Health \\
4 &Village &Nation & &School &Politics &Public & &Sector &National &Developing & &Developing &Life \\
5 &Law &Peace & &Services &Activities &Executive & &Social &Life &Production & &Sector &Developing \\
6 &Honor &Relations & &Capital &Plays &Politics & &System &Right &Program & &Investment &World \\
7 &Politics &Order & &Works &Economy &Nation & &Means &Production &Service & &Construction &Education \\
8 &Peace &Right & &Construction &Education &Administration & &Politics &Economical &Services & &Plays &Society \\
9 &Mens &Plays & &Number &Economical &Homeland & &Increase &Nation &Bank & &Increase &Social \\
10 &Revolution &Public & &Studies &Trouble &Order & &Process &Liberty &Plan & &Management &Politics \\